# Auto-scaling HTCondor pools using Kubernetes compute resources

Igor Sfiligoi

University of California San Diego, La Jolla, CA, USA, isfiligoi@sdsc.edu

Thomas DeFanti

University of California San Diego, La Jolla, CA, USA, tdefanti@eng.ucsd.edu

Frank Würthwein

University of California San Diego, La Jolla, CA, USA, fkw@ucsd.edu

HTCondor has been very successful in managing globally distributed, pleasantly parallel scientific workloads, especially as part of the Open Science Grid. HTCondor system design makes it ideal for integrating compute resources provisioned from anywhere, but it has very limited native support for autonomously provisioning resources managed by other solutions. This work presents a solution that allows for autonomous, demand-driven provisioning of Kubernetes-managed resources. A high-level overview of the employed architectures is presented, paired with the description of the setups used in both on-prem and Cloud deployments in support of several Open Science Grid communities. The experience suggests that the described solution should be generally suitable for contributing Kubernetes-based resources to existing HTCondor pools.

## 1 INTRODUCTION

The HTCondor batch workload management system [1,2] has long been used to aggregate resources from many independent resource providers and is the core technology enabling the Open Science Grid (OSG) [3]. HTCondor architecture has very few hard requirements, allowing its services to operate in virtually any environment, e.g. both with and without elevated privileges, and in restricted network environments. Resource provisioning was, however, never a core competency of the HTCondor stack, delegating that aspect to other software providers.

Kubernetes [4] is a popular container-based resource management system, that is getting significant traction in both on-prem and Cloud environments. Integration of Kubernetes-managed resources into existing HTCondor pools is thus highly desirable. As hinted above, joining a Kubernetes-managed resource, aka a Pod, to an existing HTCondor pool is relatively easy, and has indeed been happening on the Pacific Research Platform's (PRP) [5] Kubernetes cluster in opportunistic mode since 2019 [6]. This work focuses on adding a proper demand-driven, auto-scaling provisioning mechanism that is suitable for mainstream use.

Kubernetes does natively support the notion of auto-scaling a compute workload [7], which in our use-case represents HTCondor execute services that will join an existing HTCondor pool. The default auto-scaling is, however, optimized for uniform stateless services, while user jobs managed by the HTCondor execute services both possess state and can be heterogeneous. We thus decided to implement a dedicated auto-scaling solution to achieve better resource utilization. The code is open source and available in GitHub [8]. Section 2 provides an overview of the provisioning logic, while Section 3 provides some implementation details. Section 4 describes two operational modes employed to integrate Kubernetes resources into globally distributed HTCondor pools. Section 5 describes our experience in using this mechanism in preemptible environments and Section 6 explains how it interacts with Cloud auto-provisioning mechanisms.

## 1.1 Related work

Opportunistic, or backfill integration of Kubernetes resources into HTCondor pools has been used by several OSG communities when dealing with on-prem deployments, like the PRP and the Scalable Systems Laboratory [9]. It is very easy to setup and works really well when integrating with a single HTCondor pool. However, it is not suitable for sharing resources between multiple communities, especially when some of them have only intermittent compute needs.

The most widely used dynamic HTCondor resource provisioning solution is currently GlideinWMS [10], which has been in use by OSG communities for over a decade. It specializes in provisioning Grid computing resources, i.e. compute resources managed by independent batch workload management systems, but currently has no Kubernetes integration.

Manual provisioning of resources is of course always an option and may be suitable for one-off, large-scale exercises [11]. It is however relatively human effort intensive, and thus not suitable for ongoing operations.

## 2 AUTO-SCALING PROVISIONING LOGIC

The provisioning logic used for this project is relatively simple. The provisioning service keeps track of how many HTCondor jobs need additional resources and periodically compares that with the number of Kubernetes pods waiting for resources. If not enough pods are queued, more are submitted. The pods are configured to self-terminate if no user jobs are waiting for resources, automating resource provisioning scale-down.

The provisioning service allows for filtering of HTCondor jobs based on advertise attributes, so only jobs that can run on the managed Kubernetes system are considered for auto-scaling purposes. The filter is also propagated to the HTCondor execute configuration, enforcing the desired policy for all provisioned pods. The HTCondor execute services running in Kubernetes pods can also advertise additional attributes, and those attributes can be used in the filter as well.

Given that user jobs submitted to HTCondor queues tend to be heterogeneous, the provisioning service groups together jobs with similar requirements and independently requests Kubernetes resource with matching requirements, effectively creating independent filtering groups. The grouping criteria is currently based on CPU, GPU, memory and disk requirements, but could be extended in the future.

## 3 AUTO-SCALER IMPLEMENTATION

The auto-scaling provisioning service has been implemented as a Python process and is typically run inside an unprivileged pod in the Kubernetes cluster whose resources are being provisioned. A namespaced service account token [12] is used for interfacing with the Kubernetes cluster, while the HTCondor credentials must be passed as a Kubernetes secret object. The location of the HTCondor central manager must be passed as a pod environment value.

The default configuration is usable as-is and includes a CentOS-based GPU-enabled execute container image. Most resource managers will, however, likely want to add additional attributes and modify the filters. The service uses a standard Python configparser INI file [13], that is passed to the pod as a configmap object. On top of attributes and filters, the resource manager can also change the container image used, set the priority class, tolerations and node selectors, and configure storage areas. An example INI file is available in Figure 1.

```
[DEFAULT]

k8s_domain=nrp-nautilus.io

[k8s]

tolerations_list=nautilus.io/noceph, nautilus.io/suncave

node_affinity_dict=^nautilus.io/low-power:true,gpu-type:A100|A40|V100

priority_class=opportunistic

envs_dict=USE_SINGULARITY:no,GLIDEIN_Site:SDSC-PRP
```

Figure 1: Example configuration file

The code is licensed under the BSD open source license and is available on GitHub [8].

## 4 INTEGRATION WITH PRODUCTION HTCONDOR POOLS

Integrating Kubernetes-managed resources into production HTCondor pools typically means crossing organizational boundaries. This implies that not all user jobs are able or allowed to make use of the provisioned resources. The limits could be due to policy or technical reasons, e.g. restrictions on data access. Filtering, proper attribute advertisement and proper job runtime environment thus become very important.

Given the flexibility of the HTCondor ecosystem, direct integration with a production HTCondor pool typically requires a moderate understanding of the conventions and expectations of its users. This operation mode is thus best suited for when personnel associated with HTCondor operations can themselves instantiate an auto-scaling provisioning service, e.g. by being granted access to a namespace in the target Kubernetes system. We have experience provisioning Kubernetes resources this way for two HTCondor pools, one serving the UCSD physics users [14] and one serving IceCube [15] production simulation activities.

However, when user communities do not have sufficient manpower or expertise to operate yet another service, the auto-scaling provisioning system can be used by a Kubernetes resource owner to create a Grid interface, that can then be integrated with established tools like OSG's GlideinWMS. In this operation mode, the Kubernetes resource owner creates a local, dedicated HTCondor pool, alongside a Grid portal, e.g. HTCondor-CE [16], which can naturally interface with it. The auto-scaling provisioning service, still operated by the resource owner, can now be significantly simplified, as most of the user community specific configuration and policy decisions are now handled at the Grid level, e.g. as part of the pilot paradigm. There is of course some overhead involved in this layered provisioning scheme but may be the only viable option for some communities. At the time of writing, the PRP operators were supporting several OSG-affiliated communities this way.

## 5 OPERATING IN PREEMPTIBLE ENVIRONMENTS

Batch scheduling is typically used to maximize the utilization of compute resources. Ideally, thus, all compute resources are always in use, and user jobs are expected to wait in queue to achieve this level of utilization. Kubernetes is, however, often used to also serve service-oriented applications, who need relatively low startup latency. The resource administrators thus must choose between keeping some resources always unused or schedule batch pods in preemptible mode.

HTCondor transparently handles resource preemption and the auto-scaling provisioning setup also gracefully deals with preemption; batch jobs being preempted will be rescheduled to another compute resource. Enabling preemption thus likely results in higher science output, as at least some user jobs complete on resources that would otherwise be left unused. We have extensive experience running the provisioner on the PRP in preemptible mode, by assigning a lower priority to the HTCondor execute pods, as shown in Figure 1. For example, this has allowed several OSG-affiliated science communities to make good use of 350k GPU hours of compute from the PRP's Kubernetes cluster in 2021, as shown in Figure 2, without any effect on other users and services of the same Kubernetes cluster.

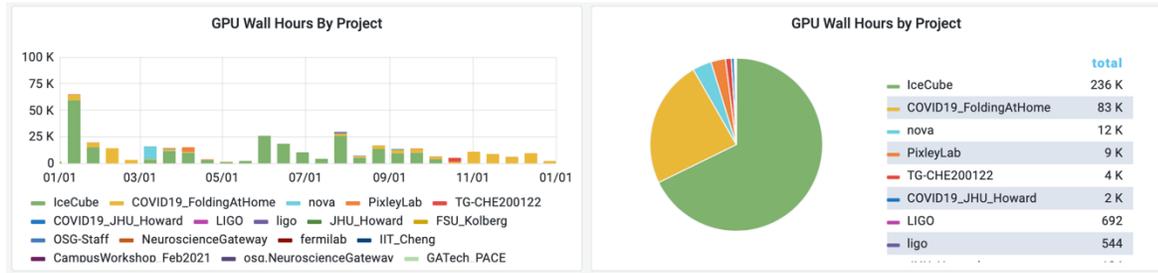

Figure 2: Screenshot of OSG GRACC GPU Payload Jobs Summary for the PRP Facility covering year 2021.
Each bin in the histogram represents 2 weeks' worth of GPU compute. (https://gracc.opensciencegrid.org/)

Preemption can also occur at Kubernetes node level, either due to hardware errors, system maintenance or hardware sharing; in all cases, HTCondor and the provisioner transparently deal with them. A particularly useful scenario is the use of spot-priced Cloud instances, which are available at significant discount but guarantee no uptime. We have been running in spot mode on the Google Kubernetes Engine (GKE) [17] for many weeks, and never experienced a problem due to preemption.

## 6 INTERACTION WITH CLOUD NODE AUTO-SCALING

Most Cloud Kubernetes deployments support the notion of node auto-scaling, i.e. autonomously changing the number of instances that identify as Kubernetes nodes based on the amount of pod request in the Kubernetes queue. This feature makes the elastic use of Cloud resources easy to use and allows the Kubernetes administrators to only pay for the amount of resources in use at any point in time.

Our auto-scaling HTCondor execute pod provisioner works seamlessly with the Cloud node auto-scaling mechanism. The number of HTCondor execute pods in Kubernetes queue varies with HTCondor-driven demand, and this in turn drives the number of Kubernetes nodes provisioned by the Cloud auto-scaler. We tested this using GKE node auto-provisioning [18], and the resulting number of provisioned Cloud resources matched nicely with the demand created by our pod provisioner, as shown in Figure 3. New Kubernetes nodes were promptly provisioned. As expected, the de-provisioning of Kubernetes nodes resulted in some unused resources, but this is unavoidable when several pods are allowed to share a single node, as was the case in our setup, since pod lifetime is dictated by user jobs, and they thus rarely terminate all at the same time. That said, we consider the observed waste in out GKE test run to be close to the minimum achievable.

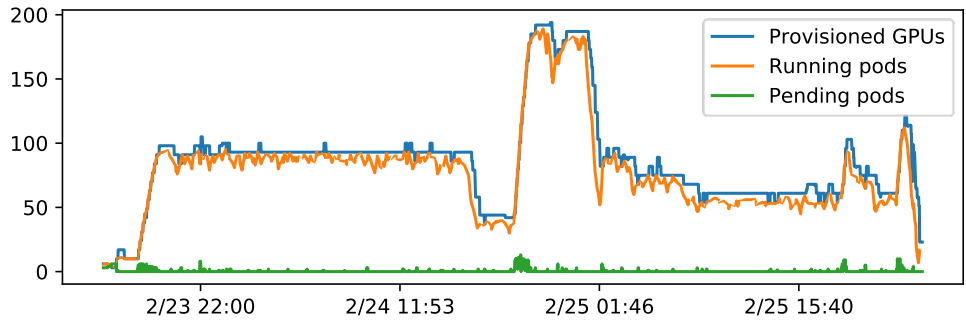

Figure 3: Monitoring snapshot of the GKE auto-scaling test run. Each provisioned Kubernetes node had 7 GPUs and each Kubernetes pod was requesting a single GPU. The pods in Kubernetes were submitted by the HTCondor-driven provisioner.

# 7 CONCLUSIONS AND FUTURE WORK

Kubernetes has become a prominent resource management solution, both on-prem and in the Clouds. Having the ability to integrate Kubernetes-managed resources into production HTCondor pools is thus highly desirable.

In this work we provide a summary overview of a new HTCondor-driven, auto-scaling provisioning tool, alongside our experience in using it to provision Kubernetes resources for several OSG-affiliated scientific communities. We integrated both on-prem and Cloud resources in HTCondor pools of those communities and are very satisfied with the operational experience so far.

The tool is available under an open source license in GitHub [8] and aims to be generic enough to be used by many user communities. The main deployments have been so far managed by our group, but we expect to expand this in the near future.

## ACKNOWLEDGMENTS

This work has been partially funded by the US National Science Foundation (NSF) Grants OAC-1826967, OAC-2030508, CNS-1925001, OAC-1841530, CNS-1730158, OAC-2112167, CNS-2100237 and CNS-2120019. All Google Kubernetes Engine costs have been covered by Google-issued credits.